\documentclass[11pt,nofootinbib,preprint]{revtex4}
\usepackage{mathrsfs}
\usepackage{graphicx}
\usepackage{amsmath}
\usepackage{amsfonts}
\usepackage{amssymb}
\usepackage{color}
\usepackage[colorlinks,linkcolor=blue]{hyperref}
\usepackage{epsfig}
\usepackage{CJK}
\usepackage{graphicx}
\usepackage{epsfig}
\usepackage{eepic}
\usepackage{bbm}
\usepackage{dcolumn}
\usepackage{bm}
\usepackage{ulem}
\usepackage{multirow}
\usepackage{slashed}

\newcommand{\omits}[1]{}

\def\bc{\begin{center}}

\def\ec{\end{center}}
\def\be{\begin{eqnarray}}
\def\ee{\end{eqnarray}}

\definecolor{dyellow}{rgb}{1.,0.8,.0}
\definecolor{myblue}{rgb}{.1,.1,.7}
\definecolor{dcyan}{rgb}{.0,.6,.6}
\definecolor{cyan}{rgb}{0.4,1.0,1.0}
\definecolor{dmagenta}{rgb}{0.6,0.0,0.6}
\definecolor{brown}{rgb}{0.6,0.2,0.}
\definecolor{darkblue}{rgb}{.0,.0,0.5}
\definecolor{darkred}{rgb}{0.75,0.0,0.0}
\definecolor{orange}{rgb}{1.,.6,.0}
\definecolor{dorange}{rgb}{0.8,.4,.0}
\definecolor{green}{rgb}{0.0,1.0,0.0}
\definecolor{darkgreen}{rgb}{0.0,0.6,0.0}
\definecolor{purple}{rgb}{.4,.0,.4}
\definecolor{lightgrey}{rgb}{0.7, 0.7, 0.7}
\definecolor{grey}{rgb}{0.4, 0.4, 0.4}

\def\R{I\!\!R}

\newcommand{\nc}{\newcommand}
\nc{\rnc}{\renewcommand} \nc{\ket}[1]{\left | \, #1 \right \rangle}
\nc{\bra}[1]{\left \langle #1 \, \right |}
\nc{\ua}{\uparrow} \nc{\da}{\downarrow}

\nc{\braket}[2]{\langle\, #1\,|\,#2\,\rangle}
\nc{\half}{\frac{1}{2}}

\nc{\prj}{\mathcal{P}} \nc{\hilb}{\mathcal{H}}
\nc{\pth}{\mathcal{C}} \nc{\inprod}[2]{\braket{#1}{#2}}
\nc{\upket}{\ket{\uparrow}} \nc{\downket}{\ket{\downarrow}}
\nc{\upbra}{\bra{\uparrow}} \nc{\downbra}{\bra{\downarrow}}

\newcommand{\ba}{\begin{aligned}}
\newcommand{\ea}{\end{aligned}}

\def\be{\begin{equation}}
\def\ee{\end{equation}}
\def\bea{\begin{eqnarray}}
\def\eea{\end{eqnarray}}

\def\l{\lambda}

\def\p{\partial}
\def\m{\mu}
\def\n{\nu}
\def\a{\alpha}

\def\>{\rangle} 
\def\<{\langle} 
\def\vev#1{\langle #1\rangle}
\def\bra#1{\left\langle #1\right|}            
\def\ket#1{\left| #1\right\rangle}

\def\Tr{\text{Tr}}

\def\vt1{\vartheta_1}

\begin{document}
\vspace{0.5cm}
 
\title{Note on $T\bar{T}$ deformed matrix models and JT supergravity  duals}

\author{Song He$^{1,2}$} \email{hesong@jlu.edu.cn }
\author{Hao Ouyang$^{1}$} \email{haoouyang@jlu.edu.cn}
\author{Yuan Sun$^{1}$} \email{sunyuan@jlu.edu.cn}
 
\affiliation{${}^1$Center for Theoretical Physics and College of Physics, Jilin University, 
Changchun 130012, China\\
${}^2$Max Planck Institute for Gravitational Physics (Albert Einstein Institute),\\
Am M\"uhlenberg 1, 14476 Golm, Germany}

\begin{abstract}
In this work we calculate the partition functions of $\mathcal{N}=1$ type 0A and 0B JT supergravity (SJT) on 2D surfaces of arbitrary genus with multiple finite cut-off boundaries, based on the $T\bar{T}$ deformed super-Schwarzian theories. In terms of SJT/matrix model duality, we compute the corresponding correlation functions in the $T\bar{T}$ deformed matrix model side by using topological recursion relations as well as the transformation properties of topological recursion relations under $T\bar{T}$ deformation. We check that the partition functions finite cut-off  0A and 0B SJT on generic 2D surfaces match the associated correlation functions in $T\bar{T}$ deformed matrix models respectively. 

  \end{abstract}


\maketitle
\tableofcontents
\section{Introduction}
In recent years, a particular kind of irrelevant deformation of field theory referred to as $T\bar{T}$ deformation \cite{Zamolodchikov:2004ce,Smirnov:2016lqw,Cavaglia:2016oda} has received much research attention. The $T\bar{T}$ deformation possesses a number of remarkable properties, e.g. integrable and solvable \cite{Smirnov:2016lqw,LeFloch:2019wlf,Jorjadze:2020ili}, helping us gain a better understanding of two-dimensional deformed quantum field theory \cite{Dubovsky:2017cnj,Cardy:2018sdv,Dubovsky:2018bmo}. These properties make the deformed theories tractable though the deformation is irrelevant. Another aspect that makes $T\bar{T}$ deformation so compelling is that in the context of  holographic duality \cite{McGough:2016lol,Kraus:2018xrn}, the $T\bar{T}$ deformed CFT was proposed to be dual to finite cut-off AdS$_3$ with positive deformation parameter \cite{McGough:2016lol}.   Nevertheless, the cut-off picture was shown valid only for pure AdS$_3$ gravity, while in the general case with the matter, the deformed CFT is dual to AdS$_3$ gravity with mixed boundary conditions \cite{Guica:2019nzm}\cite{Kraus:2021cwf}\cite{He:2021bhj}. Interestingly, the two dual pictures were consistent with each other in pure gravity. A natural question to ask is what is dual of the finite cut-off AdS gravity in other dimensions, and whether one can define the analog $T\bar{T}$ deformation in other dimensions providing such a duality. For higher-dimensional cases, it has been explored in \cite{Taylor:2018xcy}. For lower dimension, i.e., the one-dimensional case or ordinary quantum mechanic system, which is the main interest of present work,  the $T\bar{T}$ deformation was first introduced in \cite{Gross:2019ach,Gross:2019uxi}, see also recent developments \cite{Chakraborty:2020xwo,He:2021dhr,Ebert:2022gyn,Das:2022uhj, Ferko:2022iru,Ebert:2022xfh,Kruthoff:2022voq,Babaei-Aghbolagh:2022uij}.

One way to define  $T\bar{T}$ deformation in 1D is through the dimensional reduction of $T\bar{T}$ operator in 2D \cite{Gross:2019ach,Gross:2019uxi}, which is 
\be \label{TTbar1D}
2\p_t H=\frac{H^2}{4-2t H}.
\ee
Here $t$ is the deformation parameter, and one works with Hamiltonian $H$ instead of Lagrangian formalism as in 2D. It was shown that such deformation shares many crucial properties with that of  2D theory.
The deformed spectrum can be worked out explicitly which agrees with the energy of 2D black hole with a finite cut-off, and the eigenvectors remain unchanged under deformation. In this sense, the deformation is solvable. Furthermore, such deformation will preserve the supersymmetry of seed theory, which is in the same situation as in 2D cases \cite{Chang:2018dge,Jiang:2019hux,Chang:2019kiu,Coleman:2019dvf,He:2019ahx}. In addition, the partition function and correlation functions were also investigated in the deformed theory. Especially, similar to the Nambu-Goto action corresponding to the deformation of the 2D boson, the deformation of 1D QM will be a world line action. As concrete examples \cite{Gross:2019ach,Gross:2019uxi}, the deformed Schwarzian and SYK model were discussed. Among other things, a remarkable feature is the holographic dual of 1D deformed theory. The dual gravity with mixed boundary conditions at infinity can be interpreted as a finite cut-off with the Dirichlet boundary condition.

The 2D JT gravity on a disk was proposed to be dual to Schwarzian theory, which was regarded as a particular example of AdS$_2$/CFT$_1$ \cite{Maldacena:2016upp}. Since then a substantial amount of work has been focused on many aspects of JT gravity. In \cite{Saad:2019lba}, JT gravity was shown to be dual to a double scaled Hermitian random matrix model. From such a perspective, the dual of JT gravity on higher genus Riemann surface beyond disk can be explored, which is an example of ensemble average in the lower dimensional holography \cite{Schlenker:2022dyo}. The JT gravity can be viewed as a limit of a minimal string theories \cite{Seiberg:2003nm} which is known to possess random matrix description \cite{Brezin:1990rb,Moore:1991ir,Ginsparg:1993is,Mertens:2020hbs,Gregori:2021tvs}. The JT/matrix model duality has been extended in many directions. A general classification of such kind of duality was investigated in \cite{Stanford:2019vob}, including both JT and JT supergravity (SJT) cases.  Furthermore, the deformed JT gravity with a generic dilaton potential also admits matrix model description \cite{Witten:2020wvy,Turiaci:2020fjj,Forste:2021roo,Rosso:2021orf}. Interestingly, there are applications on page curves with island formula \cite{Okuyama:2021bqg} and average ensemble holography. For other related recent developments, please refer to \cite{Blommaert:2019wfy,Okuyama:2021eju,Suzuki:2021zbe,Gao:2021uro,Saad:2021uzi}. In the present work, we are interested in the JT and SJT/ matrix model duality with $T\bar{T}$ deformation\cite{Rosso:2020wir,Iliesiu:2020zld,Griguolo:2021wgy,Ebert:2022gyn,Johnson:2020heh,Johnson:2019eik,Johnson:2020mwi,Johnson:2021rsh,Johnson:2021zuo}.

In \cite{Iliesiu:2020zld}, the partition function of JT gravity with finite cut-off boundaries was in good agreement with the partition function and correlation functions of $T\bar{T}$ deformed Schwarzian theory. Subsequently, the finite cut-off JT
gravity, defined based on $T\bar{T}$ deformed Schwarzian theory, was re-visited in \cite{Griguolo:2021wgy}. The disk and trumpet partition functions were computed by the resurgence method. The result was consistent with \cite{Iliesiu:2020zld}. Meanwhile, the deformed partition functions for JT gravity with higher genus and multiple boundaries are dual to correlation functions in the matrix model, which can be derived from the Eynard-Orantin topological recursion relation in the matrix model. It is a natural question to check the  finite cut-off SJT/matrix model duality. The random matrix theory dual of SJT was introduced in \cite{Stanford:2019vob}. Since fermions are present in supersymmetric theory, there are two different ways to sum over the spin structure for fermion fields. It follows that there exist two types of SJT called type 0A and type 0B. They are dual to complex and Hermitian matrix models respectively. In the literature, the duality between SJT with certain defects and the matrix model is investigated in \cite{Rosso:2021orf}.

Inspired by the above progress, we focus on investigating the duality between the $\mathcal{N}=1$ JT supergravity (SJT) with finite cut-off and the corresponding matrix model. One can define the finite cut-off of SJT as the holographic dual of $T\bar{T}$ deformed super-Schwarzian theory. The deformed disk and trumpet partition function can be worked out by applying the techniques developed in \cite{Griguolo:2021wgy} as well as the method in \cite{Iliesiu:2020zld}. For SJT with higher genus and multiple boundaries, the partition functions can be computed using the gluing procedure  systematically. In the matrix model side, the quantities dual to gravity partition functions are the correlation functions of , for example, the resolvents, which can be computed  by using topological recursion relation. The results obtained in the matrix model match the gravity side computation in the 0A case. For 0B SJT with finite cut-off, to match the gravity side, we make use of the covariant properties of topological recursion relation under the transformation induced by $T\bar{T}$ deformation. 

The structure of this paper is organized as follows. In Section 2, we review some aspects of $\mathcal{N}=1$ JT supergravity, as well as the matrix model duality for type 0A and 0B SJT respectively. In Section 3, we investigate the finite cut-off deformed SJT and the corresponding $T\bar{T}$ deformed matrix models. The partition functions on the gravity side and the dual correlation functions on matrix model side are computed in this section. Conclusions and discussions are given in the final section. 
In the appendix, we list some relevant derivations in our analysis.

\section{Basic facts}\label{secII}
In this section, we firstly review some aspects of $\mathcal{N}=1$ JT supergravity, and the matrix model to set the notations. We mainly follow the discussions in \cite{Saad:2019lba,Stanford:2019vob,Rosso:2021orf}. The action of SJT  can be written in terms of superfields \cite{Chamseddine:1991fg,Forste:2017kwy}, see also \cite{Teitelboim:1983uy,Cangemi:1993mj,Cardenas:2018krd} 
\be\ba 
I'_{SJT}=-\frac{1}{2}\Big(i\int d^2z d^2\theta E\Phi(R_{+-}-2)+2\int du d\vartheta \Phi K\Big).
\ea\ee
Here $\Phi$ is the dilaton superfield including the dilaton field $\phi$
\be 
\Phi=\phi+\theta^\a \l_\a+i\theta\bar{\theta}F
\ee 
and the scalar curvature $R$ of the 2D manifold is contained in superfield $R_{+-}$ 
\be 
R_{+-}=A+\theta^\a\lambda_\a+i\theta\bar{\theta}C,~~C=R+\text{fermions}+\text{auxiliary fields}.
\ee
Besides the field $\phi$ and $R$, all other fields are fermions or auxiliary fields, whose precise definition  can be found in  \cite{Chamseddine:1991fg}. The second term is the Gibbons-Hawking-like term, and $(u,\vartheta)$ is the boundary superspace. The field contents in the bosonic part include metric and dilaton, and the supersymmetric partners consist of gravitino and dilatino. Integrating over auxiliary fields and turning off the fermions, the action of $I'_{\text{SJT}}$ (with proper boundary term added) reduces to the action of JT gravity
\be \label{action}
I_{\text{JT}}=-\frac{1}{2}\int_{\mathcal{M}}\sqrt{g}\phi(R+2)-\int_{\p \mathcal{M}}\sqrt{h}\phi (K-1),
\ee

An important kind of quantities in SJT relevant later is  the partition functions, i.e,   the path integral of $\mathcal{N}=1$ SJT  on a two-dimensional  surface $\mathcal{M}$  with $n$ boundaries, which can be written as \cite{Rosso:2021orf}  
\be\ba \label{SJTPFPI}
Z^{\text{SJT}} (\beta_1,...,\beta_n)=\int D{g_{\m\n}}D\phi D\Psi e^{-S_0\chi(\mathcal{M})-I_{\text{SJT}}(g_{\m\n},\phi,\Psi)}
\ea\ee
with 
\be 
\chi(\mathcal{M})=\frac{1}{2\pi}\Big(\frac{1}{2}\int_{\mathcal{M}}\sqrt{g}R+\int_{\p \mathcal{M}}\sqrt{h}K\Big).
\ee
Here $(\beta_1,...,\beta_n)$ are related to boundary conditions: the $i$-th boundary length =$\frac{\beta_i}{\epsilon}$, $\phi|_{\p \mathcal{M} }=\frac{1}{2\epsilon}$.   The fermions are denoted collectively as $\Psi$. $\chi(\mathcal{M})=2-2g-n$ is the Euler characteristic of the manifold $\mathcal{M}$ if it has $g$ handles and $n$ boundaries. We denote such manifold as $\mathcal{M}_{g,n}$. The appearance of the Euler characteristic implies that the gravity path integral admits a topological expansion in the limit $e^{-S_0}\ll 1$
\be \label{expg}
Z^{\text{SJT}} (\beta_1,...,\beta_n)=\sum_{g=0}^\infty e^{-(2-2g-n)S_0}Z^{\text{SJT}}_{g,n}(\beta_1,...,\beta_n),
\ee
where $Z^{\text{SJT}}_{g,n}(\beta_1,...,\beta_n)$ is understood as the gravity path integral on the manifold $\mathcal{M}_{g,n}$. The SJT on disk reduces to $\mathcal{N}=1$ super-Schwarzian theory containing a bosonic field and an anti-periodic (on disk) fermionic field with global OSp(2$|$1) symmetry \cite{Stanford:2017thb}. It turns out that the disk partition function $Z^{\text{SJT}}_{0,1}$ can be computed by the path integral of super-Schwarzian theory which is 1-loop exact \cite{Fu:2016vas,Stanford:2017thb,Stanford:2019vob}
\be \label{SJTPF}
Z_{\text{SJT,D}}(\beta)\equiv Z^{\text{SJT}}_{0,1}(\beta_1) =\sqrt{\frac{2}{\pi \beta}}e^{\frac{\pi^2}{\beta}}.
\ee
Beyond the disk case, the higher topological partition function can be obtained by gluing method which will be discussed in the subsequent section. In this method, we need to know the partition function on the trumpet, which similar to the disk case, results from the super-Schwarzian theory
\be \label{SJTPF1}
Z_{\text{SJT,T}}(\beta,b)=\frac{1}{\sqrt{2\pi \beta}}e^{-\frac{b^2}{4\beta}}.
\ee
Here the trumpet contains two boundaries, i.e, the geodesic and asymptotic boundaries, $b (\beta)$ is related to the length of the geodesic (asymptotic) boundary. 
From partition functions, one can get the corresponding spectral densities defined by 
\be\ba \label{Zrho2}
Z_{\text{SJT,D(T)}}(\beta )=\int_0^\infty dE e^{-\beta E}\rho_{\text{SJT,D(T)}}(E)
\ea\ee
as
\be \label{sjtDT}
\rho_{\text{SJT,D}}(E)=\frac{ \sqrt{2}}{\pi \sqrt{E}}\cosh( 2\pi\sqrt{E}),~~\rho_{\text{SJT,T}}(E,b)=\frac{\cos(b\sqrt{E})}{\sqrt{2}\pi E}.
\ee

Since SJT contains fermions which could be periodic (R) or anti-periodic (NS) for general topology, one should sum over different spin structures. One can defined the    parity $(-1)^\zeta$ for spin structures, where $(-1)^\zeta=1$ for NS spin structure, and $(-1)^\zeta=-1$ for Ramond. Then two types of SJT could be defined \cite{Rosso:2021orf}. The first one is denoted as type 0A SJT whose partition function is defined by summing over different spin structures, while the second one called type 0B SJT is defined by summing over different spin structures weighted by the parity $(-1)^\zeta$. Note for disk and trumpet, there is a unique spin structure, thus the disk and trumpet partition function is the same for both types of SJT.

In the next step, let us move on to the matrix model side. The SJT dual to an ensemble of supersymmetric  quantum mechanics (QM) \cite{Stanford:2019vob}. For SUSY QM the supercharge is related to Hamiltonian by $H=Q^2$.  There are two different matrix ensembles dual to two types of SJT respectively. For type 0A SJT including $(-1)^\zeta$ in the summation of spin structure, there is an additional   $(-1)^F$ symmetry in dual ensemble \cite{Stanford:2019vob}. Then the Hilbert space can be decomposed into two blocks with
\be
(-1)^F=\left(\begin{array}{cc}
I_N & 0 \\
0 & -I_N
\end{array} \right),~~Q=
\left(\begin{array}{cc}
0 &  M^\dagger\\
M & 0
\end{array} \right),~~H=Q^2=
\left(\begin{array}{cc}
0 & M M^\dagger\\
 M^\dagger M & 0
\end{array} \right)
\ee
where $I_N$ is $N\times N$ identity matrix, $M$ is $N\times N$ complex matrix. This implies that type 0A SJT dual to random complex matrix ensemble. The dictionary is \cite{Stanford:2019vob}
\be \label{dic0A}
Z^{\text{SJT},-}(\beta_1,...,\beta_n)=\vev{Z^-(\beta_1)...Z^-(\beta_n)}_{\text{conn.}}
\ee
with
\be \label{ZMM}
Z^-(\beta)=2\Tr e^{-\beta H}=2\Tr e^{-\beta MM^\dagger}. 
\ee
where  $Z^{\text{SJT},\pm}(\beta_1,...,\beta_n)$ represent $Z^{\text{SJT}}(\beta_1,...,\beta_n)$ in (\ref{expg}) for type 0A and 0B SJT respectively. Note the the gravity path integral on the LHS of (\ref{dic0A}) on connected manifold $\mathcal{M}_{g,n}$, and the average on the RHS is connected part of the following matrix integral \cite{Rosso:2021orf}
\be 
\vev{Z^-(\beta_1)...Z^-(\beta_n)}=\frac{1}{\mathcal{Z}}\int dM e^{-N\Tr(V MM^\dagger)}Z^-(\beta_1)...Z^-(\beta_n),~~\mathcal{Z}=\int dM e^{-N\Tr(MM^\dagger)},
\ee  
where $V(H)$ is a function of $H$ that defines the matrix model. 
It is convenient to work with the Hermitian matrix $H=MM^\dagger$ \cite{Stanford:2019vob,Rosso:2021orf}, and the corresponding spectral density and resolvent  defined with respect to $MM^\dagger$ are
\footnote{In the study of matrix models, one is usually interested in the quantities called resolvent  and spectral density In general, for a random matrix  $A$ with eigenvalues $\l_i$, the resolvent, spectral density  are defined as
\be\ba 
R(E)=\Tr\frac{1}{E-A}=\sum_{i=1}^N\frac{1}{E-\l_i},~~
\rho(E)=\sum_{i=1}^N \delta(E-\l_i).
\ea\ee  
They related to each other and the quantity $\Tr e^{-\beta A}$ through integral transformations
\be \label{RePF}
R(E)=-\int_0^\infty d\beta e^{\beta E}\Tr e^{-\beta A} ,~~~R(E)=\int^\infty_{-\infty} dE' \frac{\rho(E')}{E-E'}.
\ee}
\be \label{denreA}
\rho^-(E)=\Tr\delta(E-MM^\dagger),~~R^-(E)=\Tr\frac{1}{E-MM^\dagger}.
\ee

For type 0B SJT, there is no $(-1)^F$ symmetry. It follows that the dual matrix model is the ensemble for supercharge $Q$, which is a random Hermitian matrix. Similar to the type 0A case, the dictionary for this duality \cite{Stanford:2019vob}
\be \label{dic0B}
Z^{\text{SJT},+}(\beta_1,...,\beta_n)=\vev{Z^+(\beta_1)...Z^+(\beta_n)}_{\text{conn.}}
\ee 
with 
\be 
Z^+(\beta)=\sqrt{2}\Tr(e^{-\beta H})=\sqrt{2}\Tr(e^{-\beta Q^2}) 
\ee
and 
 \be \label{zzB}
\vev{Z^+(\beta_1)...Z^+(\beta_n)}=\frac{1}{\mathcal{Z}}\int dQ e^{-N\Tr V(Q)}Z^+(\beta_1)...Z^+(\beta_n),~~\mathcal{Z}=\int dQ e^{-N\Tr(V(Q)}.
\ee 
The corresponding 
spectral density and resolvent are then defined in terms of $Q$
\be \label{denreB}
\rho^+(E)=\Tr\delta(E-Q),~~R^+(E)=\Tr\frac{1}{E-Q}.
\ee
For later convenience, we use the notation $R^Q(E)=R^+(E)$, and define the resolvent for $H=Q^2$ 
\be \label{resH}
R^H(E)=\Tr\frac{1}{E-H}=\Tr\frac{1}{E-Q^2}.
\ee  

Let us focus on the correlation functions of matrix model, i.e., the RHS of (\ref{dic0A}) and (\ref{dic0B}). In large $N$ expansion as indicated by perturbation theory of matrix integral \cite{Saad:2019lba}  
\be \label{expansionR}
\vev{Z^\pm(\beta_1)...Z^\pm(\beta_n)}_{\text{conn.}}=\sum_{g=0}^\infty N^{2-2g-n}Z^{\pm}_{g,n}(\beta_1,...\beta_n).
\ee
Here $g$ is the genus of the double-line diagram in the matrix perturbation theory. Strictly speaking, equalities (\ref{dic0A}) and (\ref{dic0B}) hold only in the so-called double -scaled limit. In this limit, the $1/N$ is replaced by $e^{-S_0}$ \cite{Saad:2019lba} in (\ref{expansionR}).  \footnote{ To match the gravity results, one should take the continuum limit of matrix models. Naively, one takes $N \to \infty$ limit, however, in this limit, all the terms except the leading term (\ref{expansionR}) would vanish. To enhance the higher-order terms to keep all the terms in the expansion, one can take a second limit simultaneously, i.e., taking a particular coupling constant near its critical value. The critical value is determined by the configuration of spectral density in the matrix model \cite{Rosso:2021orf}. This is called double scaling limit \cite{Kazakov:1985ea,Kazakov:1989bc,DiFrancesco:1993cyw,Ginsparg:1993is}. Note the two limits are not independent of each other.   } 
This expansion is the matrix model version of (\ref{expg}). It follows that we have the relation $Z^{\text{SJT}\pm}_{g,n}(\beta_1,...,\beta_n)=Z^\pm_{g,n}(\beta_1,...,\beta_n)$,  ($Z^{\text{SJT}\pm}_{g,n}(\beta_1,...,\beta_n)$ are actually $Z^{\text{SJT} }_{g,n}(\beta_1,...,\beta_n)$ for 0A and 0B SJT in (\ref{expg}) respectively).
In the matrix model, usually $Z^\pm_{g,n}$ could be computed by the so-called topological recursion relation as we will discuss in detail in the next section.

\section{$T\bar{T}$ of SJT}
In this section, we will study the finite cut-off SJT and the dual $T\bar{T}$ deformed matrix models. The partition functions of SJT on general topologies are computed, which are shown to match the results obtained from the dual matrix model side. 
\subsection{Gravity side}
  As remarked above, by performing super-Schwarzian path integrals on a surface $\mathcal{M}_{g,n}$, the corresponding partition function $Z^\pm_{g,n}(\beta_1,...,\beta_n)$ can be obtained as \cite{Stanford:2019vob}. Firstly, let us consider the disk and trumpet topologies, since there is only one boundary condition for fermions in this geometry, the corresponding partition functions are the same for both type 0A and 0B SJT. The results are presented in (\ref{SJTPF}) and (\ref{SJTPF1}).
Comparing these results with the disk and trumpet partition functions in JT gravity   \cite{Bagrets:2016cdf,Stanford:2017thb,Saad:2019lba}  
\be \label{PFD}
Z_{\text{JT,D}}\equiv Z_{0,1}(\beta)=\frac{1}{4\sqrt{\pi}\beta^{3/2}}e^{\frac{\pi^2}{\beta}},~~ 
Z_{\text{JT,T}}(\beta,b)=\frac{1}{2\sqrt{\pi \beta}}e^{-\frac{b^2}{4\beta}}.
\ee
 the corresponding spectrum densities 
\be \label{densityzrho}
\rho_{\text{JT},D }=\frac{1}{(2\pi)^2}\sinh(2\pi\sqrt{E}),~~\rho_{\text{JT},T }=\frac{\cos(b\sqrt{E})}{2\pi \sqrt{E}},
~~
Z_{\text{JT},D(T)}(\beta)=\int_0^\infty dE \rho_{\text{JT},D(T)}e^{-\beta E}.
\ee

A key observation is that $Z_{\text{SJT,T}}$ equals to $Z_{\text{JT,T}}$ up to a numerical factor and moreover $Z_{\text{SJT,D}}$ can be reproduced by analytically continuing of $b$ in $Z_{\text{JT,T}}$, i.e.
\be \label{SJTJTT}
Z_{\text{SJT,T}}(\beta,b)=\sqrt{2}Z_{\text{ JT,T}}(\beta,b) 
\ee
and 
\be \label{SJTJTD}
Z_{\text{SJT,D}}(\beta )=2\sqrt{2}Z_{\text{ JT,T}}(\beta,b=2\pi i). 
\ee
The corresponding spectral density again is related to the trumpet density of the JT case as
\be 
\rho_{\text{SJT,D}}(E)=2\sqrt{2}\rho_{\text{ JT,T}}(E,b=2\pi i),~~\rho_{\text{SJT,T}}(E,b)= \sqrt{2}\rho_{\text{ JT,T}}(E,b ).
\ee 
For higher topologies, the partition functions can be computed by the gluing procedure as did in \cite{Stanford:2019vob}. More recently, this was generalized with defect deformation of SJT \cite{Rosso:2021orf}.

So far we have discussed the undeformed SJT gravity. When taking $T\bar{T}$ deformation into account, the deformed disk and trumpet function can be easily written down by the above observation, since (\ref{SJTJTT}) and (\ref{SJTJTD}) still hold. Based on JT results  \cite{Griguolo:2021wgy}
\be \ba \label{TTbarJTPF}
Z_{\text{JT,D}} (u,t)&=\frac{u}{\sqrt{t}}\frac{e^{-u/t}}{u^2+4\pi^2 t}I_2\Big(\frac{1}{t}\sqrt{u^2+4\pi^2 t}\Big)
 \\
Z_{\text{JT,D}} (u,t)&=
\frac{u}{\sqrt{t}}\frac{e^{-u/t}}{\sqrt{u^2-b^2t}}I_1\Big(\frac{1}{t}\sqrt{u^2-b^2 t}\Big),~~u=2\beta,
\ea\ee
where $I_1,I_2$ are modified Bessel functions. The SJT case is
\be\ba \label{ZSJTD}
Z_{\text{SJT,D}}(\beta,t)=&2\sqrt{2}Z_{\text{ JT,T}}(\beta,b=2\pi i,t)= \frac{2\sqrt{2} e^{-u/t}}{\sqrt{ t}}    \Big[\frac{u}{\sqrt{4\pi^2 t+u^2}}I_1(\frac{1}{t}\sqrt{4\pi^2 t+u^2})\Big]
\ea\ee
and
\be \ba 
Z_{\text{SJT,T}}(\beta,b,t)=\sqrt{2}Z_{\text{JT,T}}(\beta,b,t)=         \frac{ \sqrt{2} e^{-u/t}}{\sqrt{ t}}\Big[\frac{u}{\sqrt{-b^2 t+u^2}}I_1(\frac{1}{t}\sqrt{-b^2 t+u^2})\Big].
\ea\ee
Notice this results can be obtained by resurgence method as did in \cite{Griguolo:2021wgy} for the JT gravity case. Alternatively, the deformed SJT partition functions can also be produced by the integral kernel method proposed in \cite{Iliesiu:2020zld}
In this method, the deformed partition function is  \cite{Iliesiu:2020zld}
\be\ba 
Z(\beta,t)=\int_0^\infty d\beta'K(\beta,\beta')Z(\beta')
\ea\ee  
with the integral kernel
\be 
K(\beta,\beta') =\frac{\beta}{\sqrt{-t\pi}\beta'^{3/2}}e^{\frac{(\beta-\beta')^2}{t\beta'}} ,~~t<0.
\ee
This method is only well-defined for $t<0$. For $t>0$ which is related to finite cut-off geometry, one can analytically continue from $t<0$. Now apply the above method to the SJT disk partition function (\ref{SJTPF}), the deformed one is then 
\be\ba 
Z(\beta,t)=&\int_0^\infty d\beta'K(\beta,\beta')Z_{\text{SJT,D}}(\beta')=\frac{2\sqrt{2}u e^{-u/t} K_{1}(-\frac{\sqrt{u^2+4t\pi^2 }}{t})}{\sqrt{-t}\pi \sqrt{ 4\pi^2 t+u^2}},~~t<0.
\ea\ee  
Using $K_1(-z)=-K_1(z)=-i\pi I_1(z)$ and keeping only the real part \cite{Iliesiu:2020zld}, one can also obtain (\ref{ZSJTD}).
In parallel with JT case, the SJT disk partition function (\ref{ZSJTD}) can be write as an integral over a contour $C$ surrounding interval $(-\frac{1}{\sqrt{t}},\frac{1}{\sqrt{t}})$ on complex $s=\sqrt{E}$ plane, noting this interval corresponds to branch cut in deformed spectrum (\ref{TTarE}) below
\be\ba \label{ZC}
Z_{\text{SJT,D}}(\beta,t)= \int_c ds s\rho_{\text{SJT,D}}(s^2)e^{-I(s,t,u)}=\frac{ \sqrt{2}}{\pi}  \int_C ds \cosh(2\pi s)e^{-I(s,t,u)},
\ea\ee
where we follow the notation in \cite{Griguolo:2021wgy} with $u=2\beta$, and $I(s,t,u)=\frac{u}{t}(1-\sqrt{1-t s^2})$ (rewriting of (\ref{TTarE}) below). Furthermore, (\ref{ZC}) can be expressed in a form where both branches of deformed spectral density are included. To see this let us recall
that the $T\bar{T}$ deformation  defined in (\ref{TTbar1D}), 
would lead  to two branches deformed  eigenvalues 
\be \label{TTarE}
E_\pm(t)=\frac{2}{t}(1\mp \sqrt{1-tE}).
\ee
Notice that when $t>0$, the spectrum would be complex for when $E>1/t$. In addition, the undeformed eigenvalues can be recovered for the solution $E_+(t)$ as $t\to 0$. It seems that only the $E_+(t)$ makes sense as the deformed spectral. However, \cite{Griguolo:2021wgy} shows that by employing the resurgence method both branches should be included. Moreover,  when both branches are taken into consideration, the $T\bar{T}$ flow equation will be satisfied and the complex spectrum problem mentioned above will disappear \cite{Iliesiu:2020zld}. \footnote{ One problem is that the resulting deformed spectral density contains a negative part. For JT case is \cite{Griguolo:2021wgy}
\be 
\rho_{\text{JT,D}} (E,t)=\frac{1-tE/2}{4\pi^2}\sinh(2\pi\sqrt{E(1-tE/4)}).
\ee
which is negative in the range $(4/t,2/t)$.}

Explicitly, $Z_{\text{SJT,D}}(\beta,t)$ in terms of two branches spectrum is  
\be\ba
Z_{\text{SJT,D}}(\beta,t)
=&\frac{\sqrt{2}}{\pi}\int_{-1/\sqrt{t}}^{1/\sqrt{t}}ds \cosh(2\pi s)e^{-\frac{u}{t}}\Big(e^{\frac{1}{t}\sqrt{1-ts^2}}-e^{-\frac{1}{t}\sqrt{1-ts^2}}\Big)\\
=&\frac{\sqrt{2}}{\pi}\int_{0}^{1/t}d\mathcal{E} \frac{\cosh(2\pi \sqrt{\mathcal{E}})}{\sqrt{\mathcal{E}}}e^{-\frac{u}{t}}\Big(e^{\frac{1}{t}\sqrt{1-t\mathcal{E}}}-e^{-\frac{1}{t}\sqrt{1-t\mathcal{E}}}\Big)\\
=&\int_{0}^{4/t}dE\frac{\sqrt{2}}{\pi}(1-\frac{tE}{2})\frac{\cosh(2\pi \sqrt{E-tE^2/4})}{\sqrt{E-tE^2/4}}e^{-u E/2},
\ea\ee
where $s=\sqrt{\mathcal{E}}$, $E=-\frac{2}{t}(\pm\sqrt{1-t\mathcal{E}}-1)$.
From the last line the deformed spectral density of SJT can be read off, which is 
\be \label{der}
\rho_{\text{SJT,D}}(E,t)=\frac{\sqrt{2}}{\pi}\Big(1-\frac{tE}{2}\Big)\frac{\cosh(2\pi \sqrt{E-tE^2/4})}{\sqrt{E-tE^2/4}}.
\ee 
This spectral density has support on $(0,\frac{4}{t})$ and reproduces the undeformed one when $t=0$. Notice that $\rho_{\text{SJT,D}}(E,t)$ is  negative in the range $(\frac{2}{t},\frac{4}{t})$, which is similar with JT case \cite{Griguolo:2021wgy}. 

Along the same line, we can obtain the deformed spectral density for trumpet in SJT
\be  \label{dtrr}
\rho_{\text{SJT,T}}(E,b,t)= \Big(1-\frac{t E}{2}\Big)\frac{\cos(b\sqrt{E-tE^2/4})}{\sqrt{2}\pi\sqrt{E-tE^2/4}}.
\ee

Next, let us move on to consider other topologies with $g>0$ or $n>1$ in SJT. In this case, we should distinguish between two different types of SJT. The partition functions can be computed by the gluing procedure, i.e. gluing the basic building block, the trumpet partition function $Z_{\text{SJT},T}( \beta_i,b)$ for each boundary, to the supervolume $V_{g,n}^\pm(b_1,...,b_n)$, "$\pm$" to denote type 0A and 0B supervolume respectively \cite{Stanford:2019vob}.  $V_{g,n}^\pm(b_1,...,b_n)$ is an analogue to Weil-Petersson volume $V_{g,n}(b_1,...,b_n)$ in JT case. The latter is volume of moduli of space of hyperbolic Riemann surfaces of genus $g$ with $n$ geodesic boundary of length $b_1,...,b_n$ \cite{Stanford:2019vob}
\be \label{zgnsjt}
Z^\pm_{g,n}(\beta_1,...,\beta_n,t)=  \int_0^\infty b_1 db_1...\int_0^\infty b_n db_n V^\pm_{g,n}(b_1,...,b_n)Z_{\text{SJT,T}}(\beta_1,b_1,t)...Z_{\text{SJT,T}}(\beta_n,b_n,t).
\ee 
 Here we assume that
under $T\bar{T}$ deformation the gluing procedure still works as  finite cut-off JT/SJT case \cite{Gregori:2021tvs}. 
\footnote{There is a subtlety in the finite cut-off JT picture, in this case, since the boundary locates at a finite cut-off, its length of it is finite. Thus the length of geodesic boundary $b$ will greater than the boundary length and it seems that the integral range in (\ref{zgnsjt}) is ill-defined. Fortunately, as discussed in \cite{Griguolo:2021wgy}, for JT gravity the integral range would be unchanged under $T\bar{T}$ deformation. We will assume this also holds in the SJT case. As we will see, under this assumption, the results of deformed SJT will match the matrix model results.}
 And only the boundary condition is affected by the deformation, while the geodesic boundary remains the same, the super volumes would not change, thus the gluing takes the above form. 
 Intuitively, this can be understood as follows, the $T\bar{T}$ deformation will lead to a finite cut-off in the bulk as what happens in higher dimensions \cite{McGough:2016lol}. Therefore only the boundary conditions will be changed, which is relevant to $Z_{\text{SJT},D(T)}$ but not $V^\pm_{g,n}$. 

 For type 0A it takes the form as
\be\ba \label{VA12}
V^-_{g=1,n}(b_1,...b_n)=&\frac12 \frac{(-1)^n(n-1)!}{4}\\
V^-_{g=2,n}(b_1,...b_n)=&3\frac12 \frac{(-1)^n(n+1)!}{4^5}\Big[(2\pi)^2(n+2)+\sum_{i=1}^n b_i^2\Big]\\
\ea\ee
\be\ba \label{VA3}
V^-_{g=3,n}(b_1,...b_n)=\frac15 \frac{(-1)^n(n+3)!}{4^9}\Big[&(2\pi)^4(n+4)(42n+185)+84(2\pi)^2(n+4)\sum_{i=1}^nb_i^2 \\
&+25\sum_{i=1}^n b_i^4+84\sum_{i\neq j}b_i^2 b_j^2 \Big]
\ea\ee 
and $V^-_{g=0,n\geq 3}(b_1,...b_n)=0$. While for the case of type 0B, all $V^+_{g,n}$ vanish except $(g,n)=(0,2)$. For both case $V^\pm_{0,1}(b_1) $ is undefined, and  by definition $V^\pm_{0,2}(b_1,b_2)=2\delta(b_1-b_2)/b_1$. 
In the subsequent section, we will need the correlation functions of the resolvent, which following  (\ref{RePF}) is
\be\ba \label{Rpmgn}
R^{\text{SJT}\pm}_{g,n}(E_1,...,E_n,t)=(-1)^n\int_0^\infty d\beta_1...\int_0^\infty d\beta_n e^{\beta_1 E_1+...+\beta_n E_n}Z^{\text{SJT}\pm}_{g,n}(\beta_1,...,\beta_n,t).
\ea\ee
To evaluate this integral, at the first step, by substituting into (\ref{zgnsjt}), we should compute the following integral
\be 
\tilde{T}(E,b,t)=\int_0^\infty d\beta  Z_{\text{SJT,T}}(\beta_1,b_1,t)e^{\beta E}.
\ee
which is essentially computed in the JT case, since $Z_{\text{SJT,T}}$ is proportional to $Z_{\text{JT,T}}$. The result is   \cite{Griguolo:2021wgy}
\be 
\tilde{T}(E,b,t)\equiv-\sqrt{\frac{t}{2\pi}}\sum_{k=1}^{\infty}\frac{\Gamma(k+1/2)}{(1-tE/2)^k}\Big(\frac{2\sqrt{t}}{b}\Big)^k J_k\Big(\frac{b}{\sqrt{t}}\Big).
\ee
In the second step, Noticing that the supervolumes are polynomial in the power of $b^2_i$, the integral (\ref{Rpmgn}) decomposes into the following integrals  \cite{Griguolo:2021wgy}
\be\ba 
\tilde{R}_n(E,t)\equiv\int_0^\infty db   b^{2n+1}\tilde{T}(E,b,t)=-\frac{(2n+1)!(1-tE/2)}{\sqrt{2}(-E(1-tE/4))^{n+1}\sqrt{-E(1-tE/4)}}.
\ea\ee
Similarly, the integral (\ref{zgnsjt}) for deformed correlation functions of $Z(\beta)$  can be decomposed into 
\be \label{tildez}
\tilde{Z}_n(\beta,b,t)\equiv\int_0^\infty db   b^{2n+1}  Z_{\text{SJT,T}}(\beta,b,t)=\sqrt{\frac{2}{t}}n!u^{n+1}e^{-u/t}I_m\Big(\frac{u}{t}\Big).
\ee

Now we are ready to compute the deformed correlation functions of resolvent and $Z(\beta)$ from the gravity side. Using (\ref{VA12}),  (\ref{VA3}) and  (\ref{Rpmgn}), for the cylinder geometry 
\be\ba \label{R02s}
R^{\text{SJT}\pm}_{0,2}(E_1,E_2,t) =&4R^{\text{ JT}}_{0,2}(E_1,E_2,t)\\
=&\frac{t^2(1-t E_1/2)(1-tE_2/2)(tE_1^2/4+t E^2_2-E_1-E_2)}{((1-tE_1/2)^2-(1-tE_2/2)^2)^2\sqrt{-E_1(1-tE_1/4)}\sqrt{-E_2(1-tE_2/4)}}\\
&-\frac{t^2((1-tE_1/2)^2+(1-tE_2/2)^2)}{((1-tE_1/2)^2-(1-tE_2/2)^2)^2},
\ea\ee
which reduces to undeformed result $\frac{1}{\sqrt{-E_1}\sqrt{-E_2}(\sqrt{-E_1}+\sqrt{-E_2})^2}$ when $t=0$. 
And then from (\ref{tildez})
\be\ba  \label{ZSJT02}
Z^{\text{SJT}}_{0,2}(\beta_1,\beta_2,t ) =&4Z^{\text{JT}}_{0,2}(\beta_1,\beta_2,t ) =\int_0^\infty db_1db_2 b_1b_2  Z_{\text{SJT,T}}(\beta_1,b_1,t)Z_{\text{SJT,T}}(\beta_2,b_2,t)V^\pm_{0,2}(b_1,b_2)
\\
=&\frac{4u_1u_2e^{-(u_1+u_2)/t}}{t(u_1^2-u_2^2)}\Big(u_1I_0\Big(\frac{u_2}{t}\Big)I_1\Big(\frac{u_1}{t}\Big)-u_2I_0\Big(\frac{u_1}{t}\Big)I_1\Big(\frac{u_2}{t}\Big)\Big),
\ea\ee
which reproduces the undeformed result $\frac{2}{\pi}\frac{\sqrt{\beta_1\beta_2}}{\beta_1+\beta_2}$ at $t=0$. Note that for cylinder geometry, the correlators of type 0A and type 0B share the same form as presented in above. 

In what following, we turn to consider more general topologies except the cylinder and disk. For type 0B theory, all the correlators $R^{\text{SJT}+}_{g,n}$ and $Z^+_{g,n}(\beta_1,...,\beta_n,t)$ vanish
\be \label{RSJTgnB}
R^{\text{SJT}+}_{g,n}=0,~~(g,n)\neq (0,2),
\ee 
since corresponding $V^+_{g,n}=0$. For type 0A,  we list some examples ($g\leq 3$) for the correlation functions of resolvent below
\be\ba\label{RSJTgn}
R^{\text{SJT}-}_{1,n}(E_1,...,E_n,t)=& \frac12\frac{(-1)^n(n-1)!}{4}^{n/2}\prod_i^n \tilde{R}_0(E_i,t),~~\\
R^{\text{SJT}-}_{2,n}(E_1,...,E_n,t)=&2^{n/2}3\frac{(-1)^n(n+1)!}{4^5}\Big((2\pi)^2(n+2)\prod_i^n \tilde{R}_0(E_i,t)+\sum_j \tilde{R}_1(E_j,t)\prod_{i\neq j}\tilde{R}_0(E_i,t)  \Big),\\R^{\text{SJT}-}_{3,n}(E_1,...,E_n,t)=&2^{n/2}\frac15\frac{(-1)^n(n+3)!}{4^9}\Big((2\pi)^4(n+4)(42n+185)\prod_i^n \tilde{R}_0(E_i,t)\\
&+84(2\pi)^2(n+4)\sum_j \tilde{R}_1(E_j,t)\prod_{i\neq j}\tilde{R}_0(E_i,t)+25 \sum_j \tilde{R}_2(E_j,t)\prod_{i\neq j}\tilde{R}_0(E_i,t)\\
&+84\sum_{i,j,i\neq j}\tilde{R}_1(E_i,t)\tilde{R}_1(E_j,t)\prod_{k,k\neq i,j }\tilde{R}_0(E_k,t)\Big).
\ea\ee
In the next section, we will show some examples that the correlators computed in (\ref{RSJTgn})  will match the results obtained from the dual matrix model.

\subsection{Matrix model side} 
   
\subsubsection{Type 0A}\label{type0a}
In this case, as reviewed in section \ref{secII}, the dual random matrix is a complex matrix ensemble.   It follows from (\ref{ZMM}) that the spectral density of the dual matrix model is half of that of SJT (\ref{sjtDT}) 
\be \ba 
\rho^-(E )=\frac{1}{2}\rho_{\text{SJT,D}}(E )=\frac{ 1}{\pi \sqrt{2E}}\cosh( 2\pi\sqrt{E}).
\ea\ee
Under $T\bar{T}$ deformation, using (\ref{der}), the deformed spectral density reads
\be\ba
\rho^-(E,t)=\frac{1}{2}\rho_{\text{SJT,D}}(E,t)=& \frac{1}{\sqrt{2}\pi}\Big(1-\frac{tE}{2}\Big)\frac{\cosh(2\pi \sqrt{E-tE^2/4})}{\sqrt{E-tE^2/4}}.
\ea\ee

For general topological with $(g,n)$, it follows from (\ref{ZMM}) that one has 
\be  \label{RSJTRgn}
R^{\text{SJT}-}_{g,n}(E_1,...,E_n,t)(\beta,t)=2^nR^-_{g,n}(E_1,...,E_n,t).
\ee
Here $R^-_{g,n}(E_1,...,E_n,t)$ is related to $Z^-_{g,n}(E_1,...,E_n,t)$ defined in (\ref{expansionR}) by integral transformation. \footnote{It is interesting to note that $R^-_{0,2}(E_1,E_2,t)$ takes the same form for both JT and SJT cases. This is follows from $
R^{\text{SJT}-}_{0,2}(E_1,E_2,t) =4R^{\text{ JT}}_{0,2}(E_1,E_2,t)$ (see (\ref{R02s})).} To compute $R^-_{g,n}(E_1,...,E_n,t)$ in matrix model, one can employ a power tool called topological recursion relation \cite{Eynard:2004mh,Eynard:2015aea}. This recursion relation can be derived from loop equation which play the role of Ward identity in matrix model. For matrix model dual to SJT without $T\bar{T}$ deformation, the  topological recursion relation have been considered in \cite{Stanford:2019vob,Rosso:2021orf}. Below we will consider the case when $T\bar{T}$ deformation presents. The input of topological recursion relations are the deformed spectral density (or spectral curve) and $R^-_{0,2}(E_1,E_2,t)$. To be more concrete, we first define the uniformizing parameter $z$ by $E(z)=-z^2$ as in $T\bar{T}$ JT case \cite{Griguolo:2021wgy}. In terms of $z$, the recursion relation is  
\be\ba \label{rec}
W_{g,n}(z_1,...,z_n,t)= \text{Res}_{z\to 0} K(z_1,z,t) \Big[&W_{g-1,n+1}(z,-z,z_2,...,z_n,t)\\
&+\sum'_{h_1+h_2=g}\sum'_{I_1\cup I_2=J}W_{h_1,1+I_1}(z,I_1,t)W_{h_2,1+I_2}(-z,I_2,t)\Big],
\ea\ee
where the prime in the summation indicate terms containing $W_{0,1}$ are excluded. Here $J=\{z_2,...,z_n\}$. For $g=0,n=1$ the quantities $W_{g,n}$ is related to spectral density
\be \label{W01}
W_{0,1}(z,t)=i\pi \rho_{\text{MM}}(E(z))E'(z)= \sqrt{2}(2+t z^2)\frac{\cos(\pi z\sqrt{4+t z^2})}{\sqrt{4+t z^2}}.
\ee
For general $(g,n)$ $W_{g,n}$ are determined by $R^-_{g,n}$ as 
\be\ba
W_{g,n}(z_1,...,z_n,t)=\Big\{\begin{array}{l}
\Big(R^-_{0,2}(E(z_1),E(z_2),t)+\frac{1}{(E(z_1)-E(z_2))^2}\Big) E'(z_1)E'(z_2),~~ g=0,n=2,\\
R^-_{g,n}(E(z_1),...,E(z_n),t) E'(z_1)...E'(z_n),~\text{otherwise}.
\end{array}
\ea\ee
As mentioned before $R^-_{0,2}(E(z_1),E(z_2),t)$ in SJT takes the same form as in JT case, therefore $W_{0,2}(z_1,z_2,t)$ also equals to JT case, which is 
\cite{Griguolo:2021wgy}
\be\ba\label{W02JTSJT}
 W_{0,2}(z_1,z_2,t)=&\Big(R^-_{0,2}(-z_1^2,-z_2^2,t)+\frac{1}{(E(z_1)-E(z_2)^2)}\Big)E'(z_1)E'(z_2)\\
=&\frac{4(2+tz_1^2)(2+tz_2^2)}{(z_1^2-z_2^2)^(4+t(z_1^2+z_2^2))^2}\Big(2z_1z_2+\frac{4(z_1^2+z_2^2)+t(z_1^4+z_2^4)}{\sqrt{4+tz_1^2}\sqrt{4+tz_2^2}}\Big).
\ea\ee
The last quantities that remain to explain are the kernel
\be\ba
K(z_1,z,t)=&\frac{1}{2[W_{0,1}(z,t)+W_{0,1}(-z,t)]}\int_{-z}^zd z_2 W_{0,2}(z_1,z_2,t)\\
=& \frac{z(4+t z^2)(2+t z_1^2)\sec(\pi z(4+t z^2))}{\sqrt{2}(2+t z^2)(z_1^2-z^2)\sqrt{4+t z_1^2}(4+t(z^2+z_1^2))}.
\ea\ee
With the initial data $W_{0,1}$ and $W_{0,2}$ in hand, we are ready to evaluate higher topological cases via the topological recursion relations. Below we consider several examples.
\begin{itemize}
\item{$W_{1,1} $}

Applying the topological recursion relation, one have (\ref{rec}) \footnote{Here we used
\be \label{W02mm}
 W_{0,2}(z,-z,t)=\frac{(2+tz^2)^2}{z^2(4+tz^2)^2}.
\ee}
\be\ba
W_{1,1}(z_1,t)=&\text{Res}_{z\to 0} K(z_1,z,t)W_{0,2}(z,-z,t)= \frac{2+t z_1^2}{2\sqrt{2}z_1^2(4+t z_1^2)^{3/2}}.
\ea\ee
The result from SJT in (\ref{RSJTgn}) is 
\be\ba \label{G11}
R^{\text{SJT}-}_{1,1}(E  ,  t) 
=&-\frac{\sqrt{2}}{16}\frac{1-tE/2}{(-E(1-tE/4))\sqrt{-E(1-tE/4)}}=-\frac{\sqrt{2}}{4}\frac{2+tz^2}{z^3(4+tz^2)^{3/2}}.
\ea\ee
From (\ref{RSJTRgn}), we see that the result from the matrix model is in good agreement with the one obtained on gravity side
\be\ba
2  R_{1,1}(z_1,t)=2W_{1,1}(z_1,t)/(-2z_1)=& R^{\text{SJT}-}_{1,1}(z_1,t).
\ea\ee
\item{$W_{0,3}$}

The relevant topological recursion relation for $W_{0,3}$ is  
\be\ba\label{W03}
W_{0,3}(z_1,z_2,z_3,t)=\text{Res}_{z\to 0} K(z_1,z,t)&[W_{0,2}(z,z_2,t)W_{0,2}(-z,z_3,t)\\
 &+W_{0,2}(z,z_3,t)W_{0,2}(-z,z_2,t)]=0,
\ea\ee
which also matches the SJT result since $V^{-}_{0,3}=0$.
\item{$W_{1,2}$}

The relevant topological recursion relation for $W_{1,2}$ is  
\be\ba
W_{1,2}(z_1,z_2 ,t)=&
\text{Res}_{z\to 0} K(z_1,z,t) [W_{0,3}(z,-z,z_2,t)\\
&+W_{0,2}(z,z_2,t)W_{1,1}(-z,t) +W_{1,1}(z,t)W_{0,2}(-z,z_2,t)]\\
=&\frac{(2+t z_1^2)(z+t z_2^2)}{z_1^2(4+t z_1^2)^{3/2}z_2^2(4+t z_2^2)^{3/2}},
\ea\ee
where we need to use the result (\ref{W03}) derived before.
Note this result matches the SJT computation
\be\ba
R^{\text{SJT}-}_{1,2}(E_1,E_2 ,  t)  
=&\frac{(2+t z_1^2 )(2+t z_2^2 )}{z_1^3(4+t z_1^2)^{3/2}z_2^3(4+t z_2^2)^{3/2}},
\ea\ee
since
\be\ba
4 R^-_{1,2}(z_1,z_2,t)=4W_{1,2}(z_1,t)/(4z_1z_2)=& R^{\text{SJT}-}_{1,2}(z_1,z_2,t).
\ea\ee
\item{$W_{2,1 }$}

The relevant topological recursion relation for $W_{2,1}$ is  
\be\ba \label{w21mm}
W_{2,1}(z_1)=&\text{Res}_{z\to 0} K(z_1,z,t)[W_{1,2}(z,-z)+W_{1,1}(z)W_{1,1}(-z)]\\
=&-\frac{9(2+t z_1^2)(2+\pi^2z_1^2(4+t z_1^2))}{32\sqrt{2}z_1^4(4+t z_1^2)^{5/2}}.
\ea\ee
where $W_{1,2},W_{1,1}$ have been obtained before. The  gravity result is 
\be\ba 
R^{\text{SJT}-}_{2,1}(z ,  t)  = \frac{9 (2+tz^2)(2+4\pi^2 z^2+\pi^2 t z^4)}{32\sqrt{2}z^5(4+t z^2)^{5/2}}.
\ea\ee 
Thus they match each other as 
\be\ba
2 R^-_{2,1}(z_1 ,t)=2W_{2,1}(z_1,t)/(-2z_1 )=& R^{\text{SJT}-}_{2,1}(z_1, t).
\ea\ee
\end{itemize}
Note that in the above examples, the deformed $W_{g,n}$ are related to the undeformed one by   transformation
\be\label{transWgn}
 W_{g,n}(\hat{z}_1,...,\hat{z}_1,t)=W_{g,n} (z_1,...,z_n)\frac{dz_1}{d\hat{z}_1}...\frac{dz_n}{d\hat{z}_n}
\ee
with the coordinate transformation induced by $T\bar{T}$ deformation \cite{Griguolo:2021wgy}
\be \label{coortr}
\hat{E}=-\frac{2}{t}(\sqrt{1-t E}-1),
\ee
\be\label{coortr1}
\hat{E}=-\hat{z}^2,~E=-z^2~\Rightarrow z=\frac{\hat{z}}{2}\sqrt{4+t\hat{z}^2}.
\ee
Here $W_{g,n}(z_1,...,z_n)$ is undeformed one and $W_{g,n}(\hat{z}_1,...,\hat{z}_2,t)$ is deformed one. Note that formally (\ref{coortr}) is the same as the $T\bar{T}$ deformed spectral as (\ref{TTarE}).
According to \cite{Griguolo:2021wgy}, topological recursion relations   formulated in terms of differentials
\be 
w_{g,n}(z_1,...,z_n)=W_{g,n}(z_1,...,z_n)dz_1\otimes...\otimes dz_n
\ee
is covariant and takes the same form under coordinate transformation, for example, like in (\ref{coortr1}). 
\subsubsection{Type 0B}
The dual matrix model is a random Hermitian matrix for supercharge $Q$. And according to the dual dictionary (\ref{zzB}), we have  
\be\ba \label{RSJTH}
R^{\text{SJT}+}_{g,n}(E_1,...,E_n)=2^{n/2}R^{H}_{g,n}(E_1,...,E_n).
\ea\ee
where $R^{H}_{g,n}(E_1,...,E_n)$ is $n$-pt function of resolvent $R^H(E)$ defined in (\ref{resH}).

 Let us recall the undeformed case 
\be \ba\label{ZQ}
Z_{\text{SJT,D}}(\beta)=&\sqrt{2}\vev{\Tr(e^{-\beta H})}=\sqrt{2}\vev{\Tr(e^{-\beta Q^2})}=\sqrt{2}\int^{\infty}_0dx 2xe^{-\beta x^2} \rho_{ H}(x^2)\\
=&\sqrt{2}\int^{\infty}_{-\infty}dx|x|\rho_{ H}(x^2)e^{-\beta x^2}=\sqrt{2}\int^{\infty}_{-\infty}dx e^{-\beta x^2} \rho_{ Q}(x),
\ea\ee
where $x$ is eigenvalue of Hermitian matrix $Q$, while $x^2$ is the eigenvalue of system Hamiltonian $H$. According to (\ref{Zrho2}) and (\ref{sjtDT}), $\rho_H(E)$ is 
\be 
\rho_H (x^2)=\frac{1}{\sqrt{2}}\rho_{\text{SJT,D}}(x^2)=\frac{1}{\pi x}\cosh(2\pi x).
\ee
The leading spectral density for matrix ensemble $Q$ is 
\be 
\rho_Q(E)=|x|\rho_H (x^2)=\frac{\cosh 2\pi x}{\pi}.
\ee
Note that the spectral density has support on the whole real axis, in other words, it means the branch-cut of resolvent is the real axis.

In one-cut Hermitian matrix model, in general, the leading 2-point function $R^Q_{0,2}$ depends only on the position of spectral edges. For $Q$ ensemble considered here $R^Q_{0,2}$ is \cite{Stanford:2019vob}
\be \label{RQ02}
R^Q_{0,2}(x_1,x_2) =\Big\{\begin{array}{ll}
0, &\text{$x_1,x_2$ on  same side of real axis},\\
-\frac{1}{(x_1-x_2)^2},&\text{$x_1,x_2$ on  opposite sides of real axis}.
\end{array}
\ee
Using the identity
\be\label{RHQ1}
2xR^{H}(x^2)=R^Q(x)-R^Q(-x),
\ee
one obtain the follow relation for 2-pt function
\be\ba\label{H02Q}
-4x_1x_2 R^{H}_{0,2}(-x^2_1,-x^2_2)= R^Q_{0,2} (ix_1,ix_2) +R^Q_{0,2}(-ix_1,-ix_2) -R^Q_{0,2}(-ix_1,ix_2) -R^Q_{0,2}(ix_1,-ix_2). \\
\ea\ee
Substituting into (\ref{RQ02}), one obtains $R^{H}_{0,2}(-x^2_1,-x^2_2)=\frac{1}{2x_1 x_2(x_1+x_2)^2}$. In gravity side $R^{H}_{0,2}(-x^2_1,-x^2_2)$ is dual to SJT partition function on cylinder, \cite{Stanford:2019vob} showed that this is indeed the case. 

Now let us add $T\bar{T}$ deformation. The deformed spectral density is, using (\ref{der}) and (\ref{ZQ})
\be\ba\label{rhoQd}
\rho_Q(x,t)=&|x|\rho_H(x^2,t) =\frac{1}{\sqrt{2}}|x|\rho_{\text{SJT,D}}(x^2,t)\\
=&\frac{1}{\pi}\Big(1-\frac{tx^2}{2}\Big)\frac{\cosh(2\pi x\sqrt{1-tx^2/4})}{ \sqrt{1-tx^2/4}},~~x\in(-\frac{2}{\sqrt{t}},\frac{2}{\sqrt{t}} ).
\ea\ee
Next, let us consider the leading 2pt function under deformation. According to (\ref{RSJTH}), one has
\be 
R^{\text{SJT}}_{0,2}(E_1,E_2,t)=2 R^{H}_{0,2}(E_1,E_2,t),
\ee
where the LHS is computed in (\ref{R02s}) and  RHS is related to $Q$ ensemble quantity $R_{0,2}(E_1,E_2,t)$ through (\ref{H02Q}). Thus we have
\be\ba\label{r02H4Q}
  &-4x_1x_2 R^{H}_{0,2}(-x^2_1,-x^2_2,t)\\=& R^Q_{0,2} (ix_1,ix_2,t) +R^Q_{0,2}(-ix_1,-ix_2,t) -R^Q_{0,2}(-ix_1,ix_2,t) -R^Q_{0,2}(ix_1,-ix_2,t)\\
=&-2x_1x_2 \left(\frac{t^2(1-t E_1/2)(1-t E_2/2)(tE_1^2/4+tE_2^2/4-E_1-E_2)}{((1-tE_1/2)^2-(1-tE_2)^2)^2\sqrt{-E_1(1-tE_1/4)}\sqrt{-E_2(1-tE_2/4)}}\right.\\
&\left.-\frac{t^2((1-tE_1/2)^2+(1-tE_2/2)^2)}{ ((1-tE_1/2)^2-(1-tE_2/2)^2)^2}\right),~~E_1=-x_1^2,~~E_2=-x_2^2.
\ea\ee
One  simplification of above equation can be made by using the fact
\be  \label{evenr022}
R^Q_{0,2}(x_1,x_2,t)=R^Q_{0,2}(-x_1,-x_2,t).
\ee
This could result from the fact that the deformed spectral $\rho_Q(x,t)$ in (\ref{rhoQd}) is an even function in $x$. For more details please see the Appendix (\ref{app1}).
 Note that for the undeformed $R^Q_{0,2}(x_1,x_2)$ (\ref{RQ02}), this is indeed the case. \footnote{Also (\ref{evenr022}) holds for usual one-cut Hermitian matrix model, if the support of spectral density is a symmetric interval ($-a,a$), then the $R^Q_{0,2}(x_1,x_2)$ is (see (3.3.38) of \cite{Eynard:2015aea} 
\be 
 R^Q_{0,2}(x_1,x_2)=-\frac{1}{2(x_1-x_2)^2}\Big(1-\frac{x_1x_2-a^2}{\sqrt{x_1^2-a^2}\sqrt{x_2^2-a^2}}\Big).
\ee}     
One possible solution for $R^Q_{0,2}(x_1,x_2,t)$ of (\ref{r02H4Q}) can be obtained by assuming there exist a coordinate transformation between deformed and undeformed topological recursion relation. The result is presented as follows. We leave the detailed procedure for obtaining the result in Appendix \ref{appRQ02}
\be\ba\label{oneso}
 R^Q_{0,2}(x_1,x_2,t)  =\left\{\begin{array}{ll}
 R^{Q-}_{0,2}(x_1,x_2,t),&  \text{$x_1,x_2$ on   same side  of real axis,}\\
 R^{Q+}_{0,2}(x_1,x_2,t),
&\text{$x_1,x_2$ on   opposite sides of real axis,}
\end{array} \right.
\ea\ee
where $ R^{Q-}_{0,2}(x_1,x_2,t)$ is defined in Appendix \ref{appRQ02}. It ready to check this expression reduces to undeformed case (\ref{RQ02}) when $t=0$ and indeed satisfies the equation (\ref{r02H4Q}).  


Next consider the deformed higher topologies quantites $R^Q_{g,n}(E_1,...,E_n,t)$ in $Q$ ensemble.
A simplification comes from the gravity side. As discussed in previous section, the supervolumes $V^+_{g,n}$ vanish except for the case $(g,n)=(0,2)$, which leads to $R^{\text{SJT}+}_{g,n}$ with $(g,n)\neq(0,2)$ vanishing. Thus the dual $R^{H}_{g,n}$ is expected to equal zero. From (\ref{RHQ1}), then $R^Q_{g,n}=0$ with $(g,n)\neq(0,2)$. Note this results valid whether or not the $T\bar{T}$ deformation presents, since the supervolume is unchanged under $T\bar{T}$ deformation.  

In principle, the topological recursion relation would be a possible way to investigate $R^Q_{g,n}(E_1,...,E_n,t)$ , since $Q$ is a random Hermitian ensemble. However the initial data for  topological recursion relation, i.e., $R^Q_{0,2}(E_,E_2,t)$, remains to be fixed, we will adopt another way, by making use of the transformation properties of topological recursion relation under $T\bar{T}$ deformation. At the end of subsection (\ref{type0a}), the deformed and undeformed topological  recursion relations are related to each other by the coordinate transformation 
\be  \label{EhE}
\hat{E}=-\frac{2}{t}(\sqrt{1-t E}-1).
\ee
Motivated by this fact, similarly, if there exists a that there exists a coordinate transformation relates deformed and undeformed correlators in $Q$ ensemble for type 0B theory. \footnote{The $T\bar{T}$ deformation of SUSY QM system was considered in \cite{Gross:2019ach}, where the supercharge $Q$ deforms as 
\be \label{tranq1}
Q(t)=\pm \sqrt{\frac{2}{t}(1-\sqrt{1-t Q^2})},
\ee
which induced from (\ref{EhE}).} It follows by using (\ref{transWgn}) the deformed $W_{g,n}$ would  vanish since    the undeformed $W_{g,n}$ vanish, which is consistent with the prediction of the gravity side (\ref{RSJTgnB}), i.e., $R^{\text{SJT}+}_{g,n}=0$ when $(g,n)\neq(0,2)$.

\section{Conclusions and discussions} 
In the work, we study the partition functions of type 0A and 0B SJT on higher genus $g$ 2D surfaces with $n$ finite cut-off boundaries, and the dual correlation functions in the $T\bar{T}$ deformed matrix models. The disk and trumpet partition function in SJT with finite cut-off are the same for both type 0A and 0B SJT. 
For other topologies except for disk and cylinder, the deformed partition functions can be computed by the gluing procedure, which are non-zero for 0A SJT and vanish for 0B SJT. The latter case is due to the fact that the vanishing super-volume $V^+_{g,n}$ for 0B except $(g,n)=(0,2)$. In the dual matrix side, the deformed partition functions of SJT with multiple finite cut-off boundaries are corresponding to the correlation functions $R_{g,n}(E_1,...,E_n,t)$ which can be computed by employing the topological recursion relations. In the 0A SJT, we compute several $R_{g,n}(E_1,...,E_n,t)$ (or, more precisely, the quantities $W_{g,n}$) in the dual matrix model. The results from the gravity (\ref{RSJTgn}) and matrix model (\ref{W02mm})$-$(\ref{w21mm}) match each other. In addition, there is a transformation (\ref{coortr}) rule between the deformed and undeformed topological recursion relations. 

For the case of type 0B SJT, the dual matrix model is unusual. In the undeformed matrix model, i.e. $Q$ ensemble, the spectral is supported on the whole real axis. 
After taking $T\bar{T}$ into account, the deformed density in $Q$ ensemble can be worked out. Applying the covariant property of $ W_{g,n}$ and topological recursion relation in a generic Hermitian matrix model, we compute the matrix model counterparts of correlation functions in 0B SJT on surfaces with finite cut-off boundaries. In particular, as for leading 2-point function $R^{Q}_{0,2}$, we give one possible solution that satisfies the constraint (\ref{r02H4Q}) imposed by matching the SJT result. This solution is obtained by transformation properties of $R^{Q}_{0,2}$ under $T\bar{T}$ deformation, as presented in appendix \ref{appRQ02}. For other $R^{Q}_{g,n}$, to match the gravity results, as the $R^{Q}_{0,2}$ case we assume that the deformed topological recursion relations for $Q$ ensemble exist, and there exists a transformation between the deformed and un-deformed topological recursion relations. With such assumptions, we can obtain results consistent with the gravity side. 


In addition, one problem is how to explain the negative spectral density of the deformed matrix model and the finite cut-off SJT, which should not occur in the standard matrix model. This is an open question as pointed out in \cite{Griguolo:2021wgy}, which is also presented in the finite cut-off JT case. Recently, a relevant discussion on the negative spectral density appeared in \cite{Rosso:2021orf}.  This question may be related to how to define $T\bar{T}$ deformation for the matrix model. \cite{Rosso:2020wir} treated the $T\bar{T}$ deformation of the matrix model in the framework of the standard matrix model. However, the results there don't match the gravity side. Interestingly, recently, another definition of $T\bar{T}$ deformed matrix model was proposed in \cite{Ebert:2022gyn}. The $T\bar{T}$ flow equation could be satisfied in this definition. A further study on such problems would be an important direction.   

It would be interesting to apply the deformed 2D gravity/matrix model duality to Liouville gravity, or minimal string theory. It has been shown in \cite{Mertens:2020hbs} that the partition function of such theories can be calculated by the gluing procedure and they also admit dual matrix description. Therefore one may consider the $T\bar{T}$ deformation of the corresponding matrix model. 
\subsection*{Acknowledgements} 
S.H. would like to appreciate the financial support from Jilin University, Max Planck Partner group as well as the National Natural Science Foundation of China Grants (No. 12075101, No. 12047569). Y.S. is supported by the National Natural Science Foundation of China Grants (No. 12105113).
 
\appendix
\section{Equation (\ref{evenr022})}\label{app1}
In this appendix we will show that if $\rho_0(x)$ is a even function of $x$, then  $R_{0,2}(-E_1,-E_2)=R_{0,2}(E_1,E_2)$. To this end, we convert to the  2pt correlator of resolvent  to that of spectral density
\be 
\vev{R(E_1,E_2)}=\int_{-\infty}^\infty dE'_1\int_{-\infty}^\infty dE'_2\frac{\vev{\rho(E'_1,E'_2)}}{(E_1-E'_1)(E_2-E'_2)}.
\ee
Therefore the problem now is to show 
\be\label{rho2even} \vev{\rho(-E'_1,-E'_2)}=\vev{\rho(E'_1,E'_2)},\ee
which can be seen as follows. If $\rho_0(x)$ is even, the potential for matrix model $V(x)$ is even. It follows that  in the orthogonal polynomial method (for example, see \cite{Anninos:2020ccj}), the orthogonal polynomials $P_n(x)$ have definite parity, i.e., being even or odd of $x$. The 2pt correlator for spectral density for $N\times N$ random matrix is 
\be\ba \label{rPn}
&\vev{\rho(E_1,E_2)}\\=&\frac{1}{\mathcal{Z}}\int d^Nx \Delta^2(x)\sum_{i\neq j}\delta(E_1-x_i)\delta(E_2-x_j)e^{-N\sum_i V(x_i)}\\
=&\frac{2C_N^2}{\mathcal{Z}}\epsilon_{r_1...r_N}\epsilon_{s_1...s_N}\int d^NxP_{r_1}(x_1)...P_{r_N}(x_N)P_{s_1}(x_1)...P_{s_N}(x_N)\delta(E_1-x_1)\delta(E_2-x_2)e^{-N\sum_i V(x_i)}\\
=&\frac{2C_N^2}{\mathcal{Z}}\epsilon_{r_1...r_N}\epsilon_{s_1s_2r_3...r_N}P_{s_1}(E_1)P_{r_1}(E_1)P_{s_s}(E_1)P_{r_2}(E_1)e^{-N(V(y_1)+V(y_2))} h_{r_3}...h_{r_N}\\
\propto&e^{-N(V(y_1)+V(y_2))}\sum_{r_1,r_2}\frac{1}{h_{r_1}h_{r_2}}(P^2_{r_1}(E_1)P^2_{r_2}(E_2)-P_{r_1}(E_1)P_{r_2}(E_1)P_{r_1}(E_2)P_{r_2}(E_2))
\ea\ee
with 
\be
h_n\delta_{mn}=\int dx e^{-NV(x)}P_n(x)P_m(x)
,~~
\text{and} ~~
\Delta(x)=\epsilon_{r_1...r_N}P_{r_1}(x_1)...P_{r_N}(x_N).
\ee
Here $\mathcal{Z}\propto h_1...h_N$ is a normalization factor. From the last line of (\ref{rPn}), (\ref{rho2even}) follows since $P_r(x)$ have definite parity.  Note in the first line we omit the term proportional to $\delta(E_1-E_2)$, which obviously does not effect the equality (\ref{rho2even}).

\section{$R^Q_{0,2}$}\label{appRQ02}
In this appendix, we describe the procedure to obtain (\ref{oneso}).
The undeformed $R^Q_{0,2}(x_1,x_2)$ is supported on $(-a,a)$ with $a\to \infty$.  $R^Q_{0,2}(x_1,x_2)$ has a cut along $(-a,a)$ according to (3.3.37) in \cite{Eynard:2015aea}. \footnote{$\bar{W}_2(x_1,x_2)$ in \cite{Eynard:2015aea} is $R^Q_{0,2}(x_1,x_2)$ here.} Transforming to uniformized coordinate $ z_i$ by the map ((3.3.15) of that paper $a=-b$)
\be \label{unic}
x=\frac{a}{2}\Big(z+\frac{1}{z}\Big),
\ee
where $z$ is double cover of $x$ (similar to the map $x=-z^2$ appeared in the JT case). In terms of $z$. Note $R^Q_{0,2}(x_1,x_2)$ is not covariant. And the quantities being covariant in topological recursion relation is \cite{Eynard:2015aea}
\be \label{www}
W_{0,2}(z_1,z_2)=\frac{1}{(z_1-z_2)^2}=\Big(R^Q_{0,2}(x_1,x_2)+\frac{1}{(x_1-x_2)^2}\Big)\frac{dx_1}{dz_1}\frac{dx_2}{dz_2}.
\ee 
with 
\be \label{R02stan}
R^Q_{0,2}(x_1,x_2)=-\frac{1}{2(x_1-x_2)^2}\Big(1\pm\frac{x_1x_2-a^2}{\sqrt{(x_1^2-a^2}\sqrt{x_2^2-a^2}}\Big),
\ee
where there are two choices of signature since the square root is double-valued, "-" is chosen when $x_1,x_2$ are in the same sheet, and "+" when $x_1,x_2$ locate in different sheets.

To account for $T\bar{T}$ deformation we assume the coordinate transformation $z=z\hat{z}$ between deformed and undeformed  topological relations, and then $W_{0,2}$ transforms as 
\be 
 W_{0,2}(\hat{z}_1,\hat{z}_2,t)=W_{0,2} (z_1,z_2)\frac{dz_1}{d\hat{z}_1}\frac{dz_2}{d\hat{z}_2},
\ee
where the RHS is known by (\ref{www}). The LHS is related to deformed $R^Q_{0,2}$ like in undeformed case (\ref{www}) (we treat (assume) the deformed 0B as a one-cut matrix model)
\be 
 W_{0,2}(\hat{z}_1,\hat{z}_2,t)
=\Big(R^Q_{0,2}(\hat{x}_1,\hat{x}_2,t)+\frac{1}{(\hat{x}_1-\hat{x}_2)^2}\Big)\frac{d\hat{x}_1}{d\hat{z}_1}\frac{d\hat{x}_2}{d\hat{z}_2}.
\ee
From the above equations, we can obtain the deformed $R^Q_{0,2}$
\be 
R^Q_{0,2}(\hat{x}_1,\hat{x}_2,t)=\Big(R^Q_{0,2}(x_1,x_2)+\frac{1}{(x_1-x_2)^2}\Big)\frac{dx_1}{d\hat{x}_1}\frac{dx_2}{d\hat{x}_2}-\frac{1}{(\hat{x}_1-\hat{x}_2)^2}.
\ee
Here $x (\hat{x})$ is the eigenvalue of the supersymmetry charge. The  coordinate transformation between $x$ and $\hat{x}$, if it exists, is expected to be 
 is (\ref{tranq1})  (see (\ref{transWgn}))
\be  
\hat{x}(t)=  \sqrt{\frac{2}{t}(1-\sqrt{1-t x^2})},
\ee 
Now let us compute $R^Q_{0,2}(\hat{x}_1,\hat{x}_2,t)$. At first, consider $x_1,x_2$ in $R^Q_{0,2}(x_1,x_2)$ locate in different sheet (see \ref{R02stan}), the result is 
\be \ba 
&R^Q_{0,2}(\hat{x}_1,\hat{x}_2,t)\\
=&-\frac{1}{(\hat{x}_1-\hat{x}_2)^2}\\
&\pm\frac{2(t\hat{x}_2^2-2)(t\hat{x}_2^2-2)(-4a^2+\hat{x}_1\hat{x}_2\sqrt{4-t\hat{x}_1^2}\sqrt{4-t\hat{x}_2^2}\pm\sqrt{4\hat{x}_1^2-t\hat{x}_1^4-4a^2}\sqrt{4\hat{x}_2^2-t\hat{x}_2^4-4a^2})}{\sqrt{4-t\hat{x}_1^2}\sqrt{4-t\hat{x}_2^2}\sqrt{4\hat{x}_1^2-t\hat{x}_1^4-4a^2}\sqrt{4\hat{x}_2^2-t\hat{x}_2^4-4a^2})(\hat{x}_1\sqrt{4-t\hat{x}_1^2}-\hat{x}_2\sqrt{4-t\hat{x}_2^2})^2}
\ea\ee
where the $+(-)$ corresponds to the initial $x_1,x_2$ in  $R^Q_{0,2}(x_1,x_2)$ locate in the different (same) sheet. Now we take the limit $a\to 0$, $R^Q_{0,2}(\hat{x}_1,\hat{x}_2,t)$ with "-" is 
\be 
R^{Q-}_{0,2}(\hat{x}_1,\hat{x}_2,t)=-\frac{1}{(\hat{x}_1-\hat{x}_2)^2}+\frac{4(t\hat{x}_1^2-2)(t\hat{x}_2^2-2)}{\sqrt{4-t\hat{x}_1^2}\sqrt{4-t\hat{x}_2^2}(x_1\sqrt{4-t\hat{x}_1^2}-\hat{x}_2\sqrt{4-t\hat{x}_2^2})^2},
\ee
and for "+" 
\be 
R^{Q+}_{0,2}(\hat{x}_1,\hat{x}_2,t)=-\frac{1}{(\hat{x}_1-\hat{x}_2)^2},
\ee
where for convenience we renamed $R^{Q}_{0,2}(\hat{x}_1,\hat{x}_2,t)=R^{Q\pm}_{0,2}(\hat{x}_1,\hat{x}_2,t)$ in each case. Note in the  limit $t\to 0$, $R^{Q\pm}_{0,2}(\hat{x}_1,\hat{x}_2,t)$ reduces to the result (\ref{RQ02}) as it should be.

\end{document}